\documentclass{kapproc} 

\RequirePackage{graphicx}%
\RequirePackage{epsf}%
\usepackage{psfig}

\usepackage{procps} 

\upperandlowercase
\let\footnote\savefootnote
\let\footnotetext\savefootnotetext 
\let\footnoterule\savefootnoterule 
\kluwerbib 

\begin{document}


\articletitle{A Far-Ultraviolet View of Starburst Galaxies}

\chaptitlerunninghead{FUV View of Starburst Galaxies}

 \author{Claus Leitherer}
 \affil{Space Telescope Science Institute, 3700 San Martin Drive, Baltimore, MD 21218, USA}
 \email{leitherer@stsci.edu}

 \begin{abstract}
Recent observational and theoretical results on starburst galaxies related to the wavelength regime below 1200~\AA\ are discussed. The review covers stars, dust, as well as hot and cold gas. This wavelength region follows trends similar to those seen at longer wavelengths, with several notable exceptions. Even the youngest stellar populations
show a turn-over in their spectral energy distributions, and line-blanketing is much more pronounced. Furthermore, the
O~VI line allows one to probe gas at higher temperatures than possible with lines at longer wavelengths. Molecular
hydrogen lines (if detected) provide a glimpse of the cold phase. I cover the crucial wavelength regime 
below 912~\AA\ and the implications of recent attempts to detect the escaping ionizing radiation.
 \end{abstract}

\section{Background}

The astrophysically important wavelength region below $\sim$1200~\AA\ is still relatively unexplored, at least at low redshift where restframe observations must be obtained from space. Prior to the launch of FUSE (Moos et al. 2000), 
far-ultraviolet (far-UV) studies were limited to bright objects. The earliest spectral data for bright stars were obtained by Copernicus (Rogerson et al. 1973) and ORFEUS (Grewing et al. 1991), and with the UV spectrometers on the Voyager~1 and 2 spacecraft (Longo et al. 1989). Voyager~2 also succeeded in recording a far-UV spectrum of M33 (Keel 1998). HUT (Davidsen 1993) was the first instrument sensitive enough to collect spectra of faint galaxies below Ly-$\alpha$. The mission was flown on two missions and generated a rich database of far-UV spectra of actively 
star-forming and starburst galaxies. Subsequently, FUSE with its superior resolution and sensitivity fully opened the 
far-UV window to starburst galaxies. Most of this review deals with results obtained with FUSE and, to a smaller degree, 
with HUT.

\section{Stellar Populations}

\begin{figure}[]
\centerline{\psfig{file=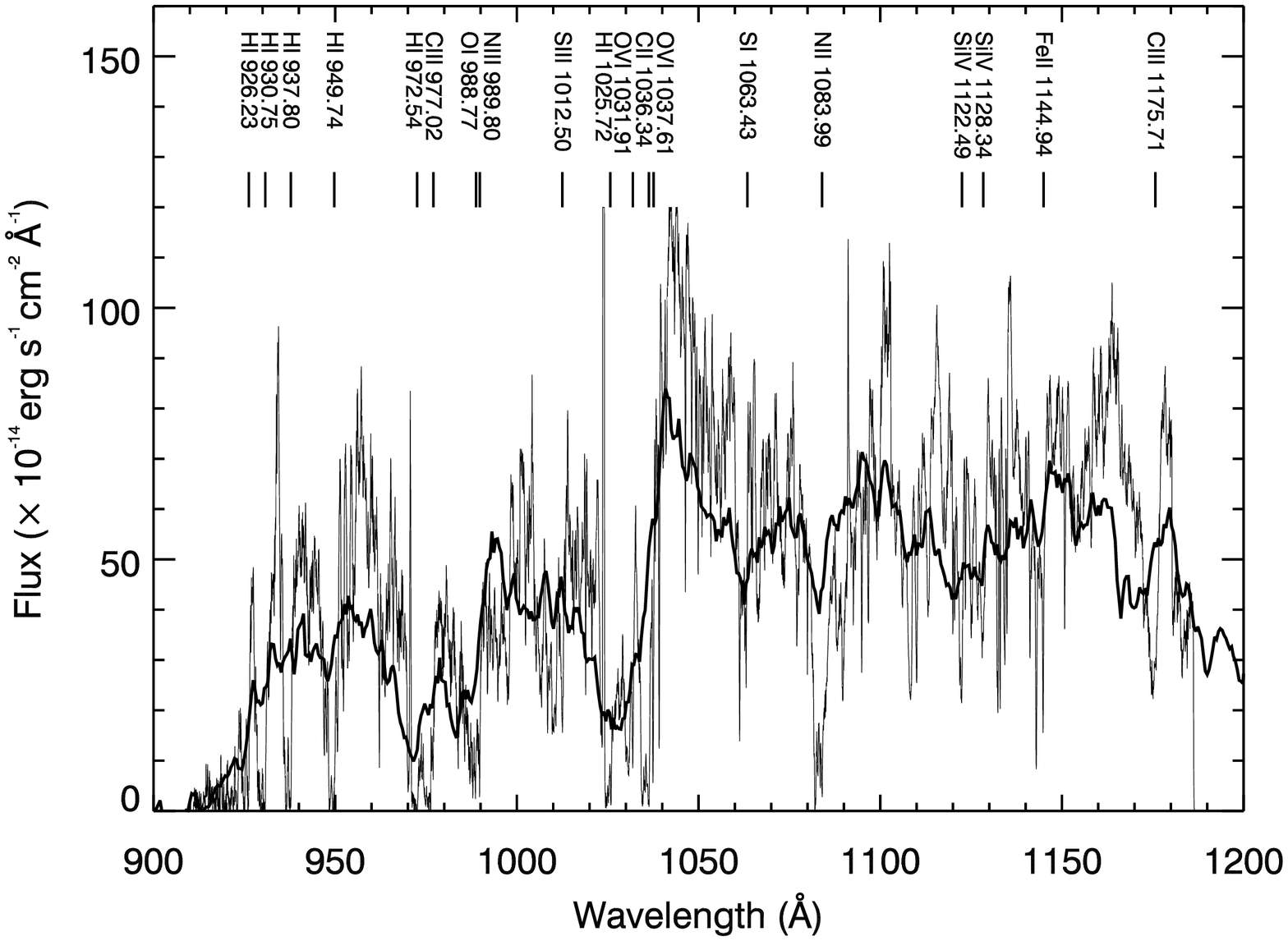,angle=90,height=2.75in}}
\caption{Spectral region between 900 and 1200~\AA\ for M83 (= NGC~5236). Major lines are labeled. Thick line: HUT; thin: FUSE (Leitherer et al. 2002).}
\end{figure}

The far-UV spectrum of the archetypal starburst galaxy M83 is reproduced in Fig.~1 (Leitherer et al. 2002). The wavelength region shown covers 900~--~1200~\AA, where blanketing is most severe. M83 has supersolar oxygen abundance. Therefore, line-blanketing effects due to stellar-wind, stellar photospheric, and interstellar lines  are particularly strong in this galaxy. The stellar features generally originate from higher ionization stages than the features found above 1200~\AA. The most prominent transition is the O~VI resonance doublet at 1032,38~\AA, which displays a spectacular 
P~Cygni profile over a broad range of spectral types. At the resolution afforded by FUSE, the blueshifted absorption component of the P Cygni profile is resolved from nearby \mbox{Ly-$\beta$} and can be distinguished from the narrow interstellar 
C~II at 1036~\AA. The (redshifted) emission component of its P Cygni profile is relatively unaffected by interstellar lines and provides additional diagnostic power. The C~III 1176 line is at the long-wavelength end of the covered spectral range and can also be observed with spectrographs optimized for wavelengths longward of Ly-$\alpha$. Surprisingly, the line has received relatively little attention in the earlier literature although it is a very good diagnostic of the properties of hot stars. C~III is not a resonance transition, and consequently does not suffer from contamination by an interstellar component. C~III, like most other stellar lines, has a pronounced metallicity dependence, either directly via opacity variations, or indirectly via metallicity-dependent stellar-wind properties.

Quantitative modeling of the stellar far-UV lines by means of evolutionary synthesis was done by Robert et al. (2003). In Fig.~2 I show an evolutionary sequence based on an empirical FUSE library of hot stars (Pellerin et al. 2002). The computed spectra are a good match to the M83 spectrum in Fig.~1. The C~III 1176 line is an excellent age diagnostic mirroring the behavior of the well-studied Si~IV 1400 line: when luminous supergiants appear around 3 Myr, wind recombination raises the emission flux (Leitherer et al. 2001). The O~VI line, in contrast, is largely decoupled from stellar parameters over a wide range of ages. This line is powered by shock heating and remains constant for stellar temperatures above 
$\sim$30,000~K. Combining lines with different optical depths, excitation, and ionization parameters allows age and metallicity estimates from far-UV spectra analogous to methods calibrated at longer wavelengths (e.g., Keel et al. 2004).

\begin{figure}[]
\centerline{\psfig{file=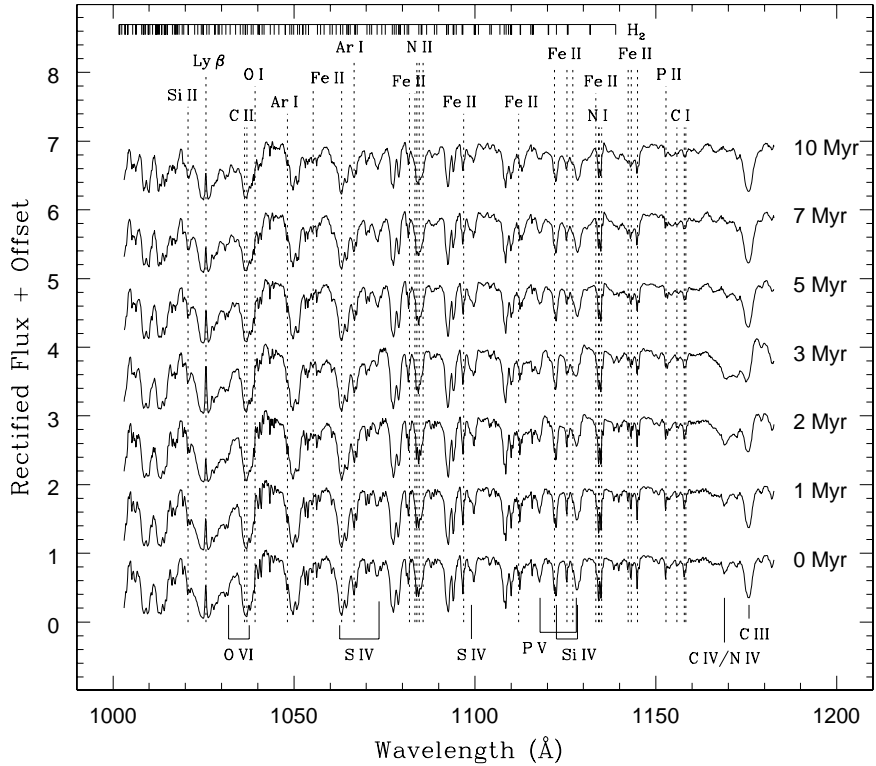,height=4in}}
\caption{Evolution of a synthetic far-UV spectrum following a Salpeter IMF between 0 and 10~Myr. Stellar features (bottom), H$_2$ lines (top), and other interstellar absorptions (top; vertical dotted lines) are labeled. The emission line at 1026~\AA\ is geocoronal (Robert et al. 2003).}
\end{figure}

Apart from their sensitivity to metallicity, the far-UV {\em lines} are affected the well-known age vs. initial mass function (IMF) degeneracy. In the absence of additional constraints, age and IMF can always be traded. This applies to the far-UV {\em continuum} as well, which in addition suffers from an age-reddening degeneracy.
In contrast to wavelengths above 1200~\AA, the intrinsic stellar spectra below 1200~\AA\ are outside the Rayleigh-Jeans regime, and age effects are no longer negligible for the continuum slope generated by an instantaneous population. Alternatively, for a population of continuously forming stars, the region between 912 and 1200~\AA\ becomes even less age sensitive to population variations than the near-UV, because an equilibrium between star formation and stellar death is reached earlier in time (see Leitherer et al. 1999). Ages and star-formation rates in starburst galaxies derived from far-UV spectra are consistent with the results from longer wavelengths.

\section{Dust Obscuration}

If the age and IMF can be constrained independently, the observed far-UV spectral energy distribution is mostly a measure of the dust attenuation. The continua of star-forming galaxies are known to obey a well-defined average obscuration curve above 1200~\AA\ (Calzetti 2001). The curve accounts for the total absorption and encompasses the net effects of dust/star geometry, absorption, scattering, and grain-size distribution. Extension of this curve down to the Lyman limit using HUT and FUSE observations of starburst galaxies was done by Leitherer et al. (2002) and Buat et al. (2002), respectively. Their results are compared to stellar data and to theoretical predictions in Fig.~3. The reddening curve of Sasseen et al. (2002) applies to individual stars; it is significantly steeper than the curves derived for galaxies. The physical interpretation of the ``grayer'' starburst reddening curve is a non-uniform attenuation. Most of the stars are totally hidden from view, and the observed UV light is emitted by those few stars which happen to have low attenuation. This effect becomes progressively more important for shorter and shorter wavelengths. The implication is that far-UV observations sample only the tip of the iceberg and could be severely biased. For instance, if there were an age- or IMF-dependence of the reddening, even the interpretation of spectral {\em lines} in the far-UV would be compromised with the assumption of a simple foreground dust screen. Circumstantial evidence for this effect to play a role has been presented by Chandar et al. (2004).

\begin{figure}[]
\centerline{\psfig{file=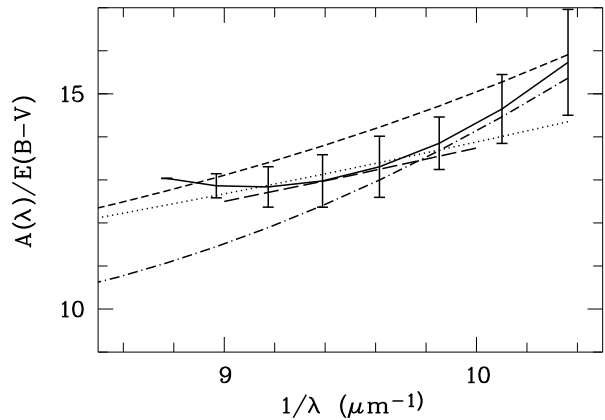,angle=-90,height=2.3in}}
\caption{Comparison of attenuation curves. Solid: Buat et al. (2002); dotted: Leitherer et al. (2002); 
short-dashed: Calzetti et al. (2000); dot-dashed: Sasseen et al. (2002); long-dashed: model of Witt \& Gordon (2000) for a shell distribution, a clumpy dust and an optical depth in the V band equal to $\tau_{\rm V} = 1.5$ (Buat et al. 2002).}
\end{figure}

Witt \& Gordon (2000) found that the empirical starburst attenuation law is most closely reproduced by a clumpy shell model with SMC-like dust and a dust column density equivalent to $\tau_{\rm V} = 1.5$. This corresponds to a far-UV attenuation correction factor of order 10 (Fig.~3).

\section{Gas: Hot and Cold Phases}

The far-UV spectral region includes line transitions that are sensitive to both the very hot ($\sim$10$^5$~K) and very cold ($\sim$10$^2$~K) interstellar material. Coronal gas with temperatures of several 100,000~K can be probed with the O~VI line whose corresponding ionization potential is 114~eV. On the other hand, the rotational/vibrational transitions of H$_2$ trace cool molecular gas.

The combined effects of multiple stellar winds and 
supernovae are capable of heating the interstellar medium (ISM) and initiating
large-scale outflows. Such outflows are a 
significant sink for the gas reservoir. They have
been known for some time from optical and X-ray imagery and have recently been analyzed with {\em absorption}-line
spectroscopy (Heckman 2005). Heckman et al. (2001) obtained FUSE far-UV 
spectra of the nearby dwarf starburst NGC~1705, probing the coronal ($10^5 - 10^6$~K) and 
the warm ($10^4$~K) phases of the outflow. The kinematics
of the warm gas are compatible with a simple model of the
adiabatic expansion of a superbubble driven by the supernovae
in the starburst. Radiative losses are negligible so that the
outflow may remain pressurized over a characteristic flow
time scale of $10^8$ to $10^9$~yr, as estimated from the
size and velocity. The same conclusion was reached for M82 by Hoopes et al. (2003)
who used FUSE to search for O VI {\em emission} in the starburst superwind of M82. 
No O VI emission was detected at any of the
pointings. These observations limit
the energy lost through radiative cooling of coronal phase gas to  the same
magnitude as that lost in the hot phase through X-ray emission, which has been shown to be small.

The total mass transported out of the starburst region via
galactic superwinds is hard to constrain, given the uncertain
ionization corrections and the strength of the observable spectral
lines. Attempts were made by  Johnson et al. (2000) and Pettini et al.
(2000) for the nearby dwarf starburst galaxy He~2-10 and the luminous 
Lyman-break galaxy MS1512-cB58, respectively.
In both cases the mass-loss rate of the ISM is rather similar to the
star-formation rate. Taken at face value, this
suggests that the available gas reservoir will not only be 
depleted by the star formation process but, more importantly, by removal
of interstellar material. Starbursts may determine their own fate by
their prodigious release of kinetic energy into the ISM.

The spectral range of FUSE includes numerous
transitions of molecular hydrogen, making it possible to study H$_2$ in diffuse interstellar environments directly
through absorption measurements, rather than relying on the indirect CO technique. Hoopes et al. (2004) searched 
for H$_2$ absorption in the five starburst galaxies NGC~1705,
NGC~3310, NGC~4214, M83, and NGC~5253. Weak H$_2$ absorption was detected 
in M83 and NGC 5253, and upper limits on the H$_2$ column density were derived in the other three galaxies. The
upper limits on the mass of molecular gas are orders of magnitude lower than the
H$_2$ mass inferred from CO emission measurements. The interpretation is that almost all of the H$_2$ is confined to
clouds with column densities in excess of $10^{20}$~cm$^{-2}$ that are
opaque to far-UV light and cannot be detected in the FUSE spectra. The far-UV
continuum seen with FUSE originates from sightlines between the dense clouds, which have a covering factor $< 1$. This morphology is consistent with that of the interstellar dust, which is thought to be clumpy.
The complex observational biases related to varying extinction across the extended UV emission in the FUSE
apertures make it difficult to characterize the diffuse H$_2$ in these starburst galaxies.

\section{Gas: Transparency of the Lyman Continuum}

Star-forming galaxies are the dominant contributor to the non-ionizing UV radiation field in the universe. Are they a significant component of the ionizing background as well? Simple arguments might suggest otherwise. An H~I
column density of $\sim$1$\times 10^{18}$~cm$^{-2}$ is sufficient to absorb
essentially all the ionizing radiation. Since the measured extinctions imply
column densities that are three or four orders of magnitude higher than this,
it might appear that essentially no ionizing radiation can escape. However, the porosity of the ISM seen in the non-ionizing continuum ($\lambda > 912$~\AA) could very well
extend below the Lyman edge and may dominate the shape of the emergent spectrum.
The situation is sufficiently complex that the only
way to determine the escape fraction $f_{{\rm esc}}$ of the ionizing radiation
is via a direct measurement. 

\begin{table}[]
\caption
{Determinations of the Lyman escape fraction in galaxies.}
\begin{tabular*}{\textwidth}{@{\extracolsep{\fill}}lccc}
\sphline
Reference& Instrument & Objects &$f_{{\rm esc}}$\cr
\sphline
Leitherer et al. (1995)       &   HUT      &  4 galaxies; $z \simeq 0$ & $<3$\%\cr
Hurwitz et al. (1997)         &   HUT      &  4 galaxies; $z \simeq 0$ & $<19$\%\cr
Deharveng et al. (1997)       &H$\alpha$/UV&local luminosity function & $<1$\%\cr
Deharveng et al. (2001)       &   FUSE     &  Mrk 54 $z \simeq 0$      & $<5$\%\cr
Giallongo et al. (2002)       &   FORS2    &  $z = 2.96, 3.32$         & $<16$\%\cr
Fern\'andez-Soto et al. (2003)&  WFPC2     &  HDF; $1.9 < z < 3.5 $    & $<4$\%\cr
Malkan et al. (2003)          &   STIS     &   $1.1 < z < 1.4 $        & $<1$\%\cr
Steidel et al. (2001)         &Keck/LRIS   &29 galaxies; $z \simeq 3.4$&{\bf  $\sim$100\%}\cr
\sphline
\end{tabular*}
\end{table}

Attempts to measure $f_{{\rm esc}}$ fall into two categories: observations of local galaxies with a far-UV detector, or measurements using galaxies at cosmological redshift, which are accessible from the ground with 8-m class telescopes. Either technique has its advantages and disadvantages. The ``local'' approach faces the obvious challenge of extreme UV observations, whereas the ``distant'' measurement is affected by the radiative transfer in the intergalactic medium. In addition, a somewhat less direct method is to determine the Lyman continuum opacity from a comparison of the H$\alpha$ and UV luminosity functions in the local universe. Table~1 gives a summary of recent results. Except for the last entry in this table, all quoted studies find more or less stringent upper limits on $f_{{\rm esc}}$ both in the low- and
high-redshift universe. The ISM in the observed galaxies is highly opaque, and very little stellar ionizing radiation leaks out.

Steidel et al. (2001) detected significant Lyman-continuum flux in the composite spectrum of 29 Lyman-break galaxies with redshifts $z = 3.40 \pm 0.09$.  If the inferred escaping Lyman-continuum radiation is typical of galaxies at $z \approx 3$,  then these galaxies produce about 5 times more H-ionizing photons per unit comoving volume than quasars at this redshift, with the obvious cosmological implications. Haehnelt et al. (2001) fitted the composite spectrum by a standard stellar population and no intrinsic H I opacity. Therefore, $f_{{\rm esc}}$ must be close to 100\% for the observed sample. Confirmation or rejection of this striking result will be a major objective of observational cosmology.

\end{document}